\documentclass{aa}
\usepackage[varg]{txfonts}
\usepackage{graphicx}
\usepackage{siunitx}
\usepackage{dcolumn}
\usepackage{xcolor}
\usepackage[normalem]{ulem}
\usepackage{subfig}

\providecommand\half{\ensuremath{\frac{1}{2}}}
\providecommand\thalf{\ensuremath{\frac{3}{2}}}
\newcolumntype{d}{D{.}{.}{-1}}
\usepackage{soul}

\allowdisplaybreaks
\graphicspath{{Figures/}}

\begin{document}

\title{On the properties of 0.11 keV to 344 MeV ion spectra in the inner heliosheath using regularized $\kappa$-distributions}

\author{K. Scherer\inst{\ref{inst1},\ref{inst2}}
  \and K. Dialynas  \inst{\ref{inst3}} 
  \and H. Fichtner\inst{\ref{inst1},\ref{inst2}}
  \and A. Galli \inst{\ref{inst4}}
  \and E. Roussos \inst{\ref{inst5}}}

\institute{Institut f\"ur Theoretische Physik, Lehrstuhl IV:
  Plasma-Astroteilchenphysik, Ruhr-Universit\"at Bochum, D-44780 Bochum,
  Germany, \email{kls@tp4.rub.de}\label{inst1}
  \and
Research Department, Plasmas with Complex Interactions,
  Ruhr-Universit\"at Bochum, 44780 Bochum, Germany\label{inst2}
  \and
  Office of Space Research \& Technology, Academy of Athens, 10679 Athens, Greece,~\email{kdialynas@phys.uoa.gr} \label{inst3}
  \and
  Physics Institute, University of Bern, Bern, Switzerland \label{inst4}
  \and
  Max Planck Institute for Solar System Research, Justus-von-Liebig-Weg 3, D-37077 Goettingen, Germany \label{inst5}}

\date{}

\date{}
\abstract{The shape of the ion energy spectra plays a critical role toward determining the ion energetics, the acceleration mechanisms and the possible sources of different plasma and suprathermal ion populations. 
The determination of the exact shape of the total particle spectrum, 
provide the necessary means to address the inner heliosheath dynamics.
Apart from various modeling efforts, 
a direct fit to the measured ion spectra for an extended energy range of $\sim$0.11 to 344 MeV has not been performed to date.}
{We use an extended set of combined 0.11-55 keV remotely sensed ENA measurements from the Interstellar Boundary Explorer (IBEX-Lo and IBEX-Hi) and Cassini/Ion and Neutral Camera (INCA), converted to protons, together with $\sim$28 keV to 344 MeV in-situ ion measurements from the Low Energy Charged Particle (LECP) and Cosmic Ray Subsystem (CRS) experiments on Voyager 2, over the declining phase of Solar Cyle 23 (SC23) and ascending phase Solar Cylce 24 (SC24)  (2009-2016) to study the characteristics of the particle energy spectrum.
}
{We fit the 0.11 keV to 344 MeV composite ion spectra with a set of regularized isotropic $\kappa$-distribution functions (RKDs)%
allowing the determination of the macroscopic physical properties.}
{We demonstrate that the 2009-2012 composite spectrum that corresponds to the declining phase of SC23 is well fitted by three different RKDs, while the 2013-2016 spectrum, associated with the rise of SC24, can only be approximated with six different $\kappa$-distribution functions.}
{Our results are generally consistent with shock accelerated particles that undergo additional acceleration inside the inner heliosheath. We identify a low energy transmitted population of particles, a suprathermal reflected population and a very high energy component that is modulated by galactic cosmic rays. The 2013-2016 time period is most likely associated with a mixture of particles from 
SC23 and 
SC24, which is reflected by the need to employ six RDKs.
}


\maketitle

\section{Introduction}\label{sec:introduction}

The solar wind (SW) interacts directly with the Very Local Interstellar Medium (VLISM), forming the heliosphere \citep{parker1961}. The neutral interstellar atoms that enter the heliosphere become ionized by various processes, i.e. photo-ionization, electron-impact ionization and charge-exchange \citep[see e.g.][]{Scherer-etal-2014}, and get "picked up" by the SW where they are accelerated to higher energies, forming a unique distribution of ions with a cut-off at roughly twice the solar wind bulk speed, called "Pickup ions" (PUI). For example, observations from the New Horizons spacecraft at $\sim$38 AU \citep{mccomas2017} showed that the PUI distribution is heated in the frame of the SW with increasing distance, before reaching the inner boundary of our solar bubble, called Termination Shock (TS). The region between the TS and the heliopause (HP) is traditionally been refereed to as "inner heliosheath" (IHS), and the region upstream at the HP is called "outer heliosheath" (OHS). Recent studies \citep[e.g.][]{dialynas2021, mccomas2020} call these regions as "heliosheath" (HS) and "Very Local Interstellar Medium" (VLISM; \cite{holzer1989,zank_2015}), respectively. The existence of an OHS (and consequently an IHS) does not depend on the existence of a bow shock/wave in front of the heliosphere, a subject that is debated in the literature \citep[e.g.][]{mccomas2012,scherer2014}, because the pristine ISM has to flow around the obstacle heliopause, and will eventually build a hydrogen wall and influence the transport of low energetic cosmic rays. We select to use the terms IHS and OHS in the present study.


The TS crossings of the Voyager 1 (V1) and Voyager 2 (V2) missions in 2004 ($\sim$94 AU; \cite{decker2005, stone2005}) and 2007 ($\sim$84 AU; \cite{decker2008, stone2008}), respectively, led to the discovery of the reservoir of ions and electrons that constitute the IHS, where the particle pressure exceeds substantially the magnetic field pressure ($\beta$=P$_{\mathrm{particle}}$/P$_{MAG}>>$1; \cite{decker2015, dialynas2019, Dialynas-etal-2020}). The IHS extends out to the HP, which serves as the interface between the SW plasma and the VLISM, that was measured in-situ by V1 in 2012 ($\sim$122 AU; \cite{burlaga2013, gurnett2013, krimigis2013, stone2013}) and by V2 in 2019 ($\sim$119 AU; \cite{burlaga2019, gurnett2019, krimigis2019, richardson2019, stone2019}, revealing the dynamic nature of the boundary, that involves an increase in Galactic Cosmic Rays (GCR), magnetic field and temperature, a simultaneous decrease in suprathermal particles, invading of interstellar flux tubes into the IHS, pointing to the possibility of flux tube interchange instability at the boundary \citep[e.g.][]{krimigis2013, florinski2015} and an outflow of suprathermal ions upstream at the HP out to $\sim$28 AU, originating from inside the IHS \citep{dialynas2021}. The activity upstream at the HP is summarized in \cite{gurnett2021}, whereas a summary of the effects of the solar activity to the particle dynamics in interplanetary space, the IHS and the OHS is presented in \cite{hill2020}.

The crossings of both Voyagers from the TS proved that the shocked thermal plasma in the region downstream of the TS remained supersonic. The bulk of the upstream energy density of the solar wind went into heating the PUIs, whereas a substantial amount of $\sim$15\% was transferred to suprathermal ($>$28 keV) particles \citep{richardson2008, decker2008}. The potential role played by electrons was studied by \citet{Chalov-Fahr-2013}. Moreover, despite previous expectations \citep[e.g.][]{pesses1981} that date back to the 80's, the $\sim$ 10-100 MeV intensities at both V1 and V2 did not peak at the TS, pointing to the understanding that the Anomalous Cosmic Rays (ACRs) are not accelerated at the TS, but most likely in the IHS by various mechanisms \citep[e.g.][]{lagner2006, Ferreira-etal-2007, lazarian2009, drake2010, strauss2010, fisk2013, zieger2015}. 

The contribution of remotely sensed ENA observations from the Interstellar Boundary Explorer (IBEX; \citep{mccomas2009}) and particularly the Cassini/Ion and Neutral Camera (INCA; \cite{krimigis2009}) that covers the $>$5.2 to $\sim$28 keV part of the PUI distribution that is not measured in-situ by the Voyagers, was unequaled (see also Section \ref{sec:intrumentation}). These observations further corroborate the complexity of the $\sim$10 eV to $\sim$344 MeV energy spectra inside the IHS \citep{Dialynas-etal-2020}, revealing a number of softening and/or hardening breaks that possibly correspond to acceleration mechanisms inside the IHS that are not currently described by sophisticated recent models \citep[e.g.][]{gkioulidou2022}. A recent model from \cite{Zirnstein-etal-2021} showed that a proton distribution that is consistent with the IBEX measurements, requires a ten times higher turbulence power ratio at the TS that that observed by Voyager \citep{burlaga2008}. Overall, the shape of the ion energy spectra plays a critical role toward determining the ion energetics, the acceleration mechanisms and the possible sources of different plasma and suprathermal ion populations. 

Heliospheric plasmas, like the SW or that in the IHS, involve non-thermal acceleration processes. Their distribution do not longer exhibit a Maxwellian (i.e., exponential) cutoff, but often a (decreasing) power law, which can be parameterized by  $\kappa$-distributions, introduced empirically by \cite{Olbert-1968}, and published for the first time by \cite{Vasyliunas-1968} as a global fitting model. More rigorous analyses involve a combination of multiple (anisotropic) distribution functions, including Maxwellian and $\kappa$ distributions \citep{Maksimovic-etal-2005, Stverak-etal-2008}.
The standard $\kappa$ distributions (SKDs) are a powerful tool for modeling non-thermal distributions, but have a critical limitation in defining the velocity moments of order $l$ (e.g. the macroscopic physical properties),  which diverge for low values of $\kappa < (l +1)/2$   \citep{Lazar-Fichtner-2021}. To heal this behavior \cite{Scherer-etal-2017} introduced a generalisation of the isotropic standard $\kappa$ distribution: the regularised $\kappa$ distribution, for which all velocity moments converge \citep[see also][for a more general discussion]{Scherer-etal-2020}. We shortly repeat its definition in   
in Sec.~\ref{sec:models}. Despite their limitations, $\kappa$-distributions were used to model the  evolution of pickup ions in the IHS \citep[e.g.][]{Fahr-etal-2016} and electrons \citep[e.g.][]{Fahr-etal-2017}. It is also used to model the interaction of protons with neutral hydrogen in the IHS \citep{Heerikhuisen-etal-2008, livadiotis2011}. For further readings we recommend \citet{Lazar-Fichtner-2021} and \citet{Livadiotis-2017}.

In the present study we model the 0.11 keV to 344 MeV composite ion spectra in the IHS, presented in \cite{Dialynas-etal-2020}, with regularized $\kappa$ distributions, following the method of \cite{Scherer-etal-2020}. Section \ref{sec:intrumentation} provides a brief overview of the measurements employed in our analysis, whereas our model is thoroughly explained in Section \ref{sec:models}. The results are presented in Section \ref{sec:results} and are subsequently discussed in Section \ref{sec:discussion}, in accordance with various studies published in the literature. Finally, Section \ref{sec:summary} provides a brief summary of our findings.

\section{Measurement details and instrumentation}\label{sec:intrumentation}
The energy spectra used in the present study (Fig.~\ref{fig:rkd}) are taken from \cite{Dialynas-etal-2020} and represent a combination of remotely sensed 0.11 keV to 55 MeV H$_{ENA}$ (converted to protons) from the Interstellar Boundary Explorer (IBEX) and Cassini/Ion and Neutral Camera (INCA) and $\sim$28 keV to 344 MeV \textit{in-situ} ions from the Voyager 2 (V2) mission, over the time period from the beginning of 2009 to the end of 2016. These energy spectra can be thought as representative of the IHS conditions during the declining phase of solar cycle 23 (SC23) and the rise of SC24 in the direction of V2. 

The Energetic Neutral Atoms (ENAs), created by the proton-neutral gas interactions through the charge-exchange (CE) process \citep[e.g.][]{lindsay2005, Fahr-etal-2007, Scherer-etal-2014}, are remotely detected by IBEX (located at $\sim$1 AU) and Cassini/INCA (at $\sim$10 AU; end of mission 15-Sep.-2017) that provide full sky images of the heliosphere \citep[e.g.][]{mccomas2009, krimigis2009}. 

Specifically, the IBEX mission includes two single-pixel ENA cameras, namely the IBEX-Lo \citep{fuselier2009} taking measurements from $\sim$10 eV to $\sim$2 keV \citep{galli2016} (a new data release can be found in \cite{galli2022b}), and the IBEX-Hi \citep{funsten2009} that measures ENAs from $\sim$520 eV to $\sim$6 keV
\citep[e.g.][]{mccomas2014}. The INCA instrument on Cassini \citep{krimigis2009}, part of the Magnetospheric Imaging Instrument (MIMI; \cite{krimigis2004}) utilized a large geometry factor (G$\sim$2.4 cm$^2$ sr) and a broad field of view (FOV; 90$^{\circ}\times$120$^{\circ}$), possessing a high sensitivity to detect very low ENA intensity events in the heliosphere and image large parts of the sky sphere over the energy range of $\sim$5.2 to 55 keV \citep[e.g.][]{dialynas2017}.

The first images of the IBEX mission at $<$6 keV showed the existence of a bright and narrow stripe of ENA emissions that cuts through the region between the V1 and V2 positions toward the upwind hemisphere and roughly encircles the global heliosphere \citep{mccomas2009, schwadron2009}. This unexpected feature, called "ribbon" is thought to lie beyond the HP and formed through a secondary ENA process \citep{Heerikhuisen-etal-2010}. The ribbon is superimposed over the so-called globally distributed flux (GDF), a "background" ENA emission that exhibits different characteristics from the ribbon \citep{livadiotis2011, schwadron2011} and is thought to be formed from within the IHS \citep[e.g.][]{dayeh2011, schwadron2014,  mccomas2009, mccomas2017, mccomas2020}. The $<$6 keV IBEX/ENA measurements in \cite{Dialynas-etal-2020} (also shown in Fig. \ref{fig:rkd}, converted to ions) do not take contributions from the ribbon.
    
At higher energies, the $>$5.2 keV ENAs measured by Cassini/INCA have shown the existence of a "Belt" of varying ENA intensities identified as relatively wide and nearly energy-independent ENA region that wraps around the celestial sphere and two "Basins" where the ENA minima occur \citep{krimigis2009, dialynas2013}. Both these features have been proven undeniably to be of IHS origin \citep[e.g.][]{krimigis2009, krimigis2011, dialynas2013, dialynas2017, dialynas2017b, dialynas2019, Dialynas-etal-2020}.

The ENAs from 0.11 to 55 keV (from IBEX-Lo, IBEX-Hi and INCA) are sampled in the pixels enclosing the position of V2 and are converted to protons (shown in Fig.\ref{fig:rkd}) using an interstellar neutral hydrogen distribution of n$_{H}\sim$0.12 cm$^{-3}$ \citep{dialynas2019, swaczyna2020} and a line of sight (LOS) of L$_{HS}\sim$35 AU (as measured by V2; e.g. \cite{decker2008, krimigis2019}), i.e. assumed to form through CE interactions from inside the IHS. For thorough information concerning the combination of the measurements shown in Fig. \ref{fig:rkd}, the reader should refer to \cite{Dialynas-etal-2020} and references therein.

The $\sim$28 to 3.5 MeV data shown in Fig.\ref{fig:rkd} are in-situ ions from the Low Energy Charged Particle (LECP) detector on V2 (\cite{krimigis1977}), that measures the differential intensities of
ions within the 28 keV to $\sim$60 MeV/Nuc energy range, together with
an integral ion measurement $>$211 MeV \citep[e.g.][]{decker2015}, whereas the $\sim$3 to 344 MeV part of the energy spectra in Fig.\ref{fig:rkd} are taken from the Cosmic Ray Subsystem (CRS) instrument \citep{stone1977} that resolves the energy
spectra and elemental composition of all cosmic-ray nuclei
from hydrogen through iron over the energy range from
$\sim$1 to 500 MeV/nuc \citep[e.g.][]{cummings2016}.

The shape and properties of the energy spectra shown here, together with specific details about the measurement techniques and analyses concerning the IBEX, INCA, LECP and CRS capabilities and data, are thoroughly explained in \cite{Dialynas-etal-2020, dialynas2022, galli2022} and references therein.

\section{The model}\label{sec:models}
Past modeling efforts \cite[e.g.][]{zirnstein2015} used a single $\kappa$-distribution over an energy range of $\sim$0.1 to 100 keV to characterize the PUI spectrum downstream of the TS and examined the change in this proton distribution due to the charge exchange losses inside the IHS, showing that the IHS spectra do not retain their initial shape. \cite{dialynas2019} further showed that the use of a single $\kappa$-distribution downstream of the TS may fit the multi-hundred kilo-electron-volt part of the ion distribution, but it would substantially underestimate the particle fluxes in (at least) the $\sim$5.2-24 keV energy range. Clearly, the description of the proton spectra inside the IHS and -perhaps- even at the TS, especially when considering an extended energy range of $\sim$0.11 keV to 344 MeV, requires the use of multiple $\kappa$-distributions.

\begin{figure*}[t!]
  \centering
  \includegraphics[width =0.95\columnwidth]{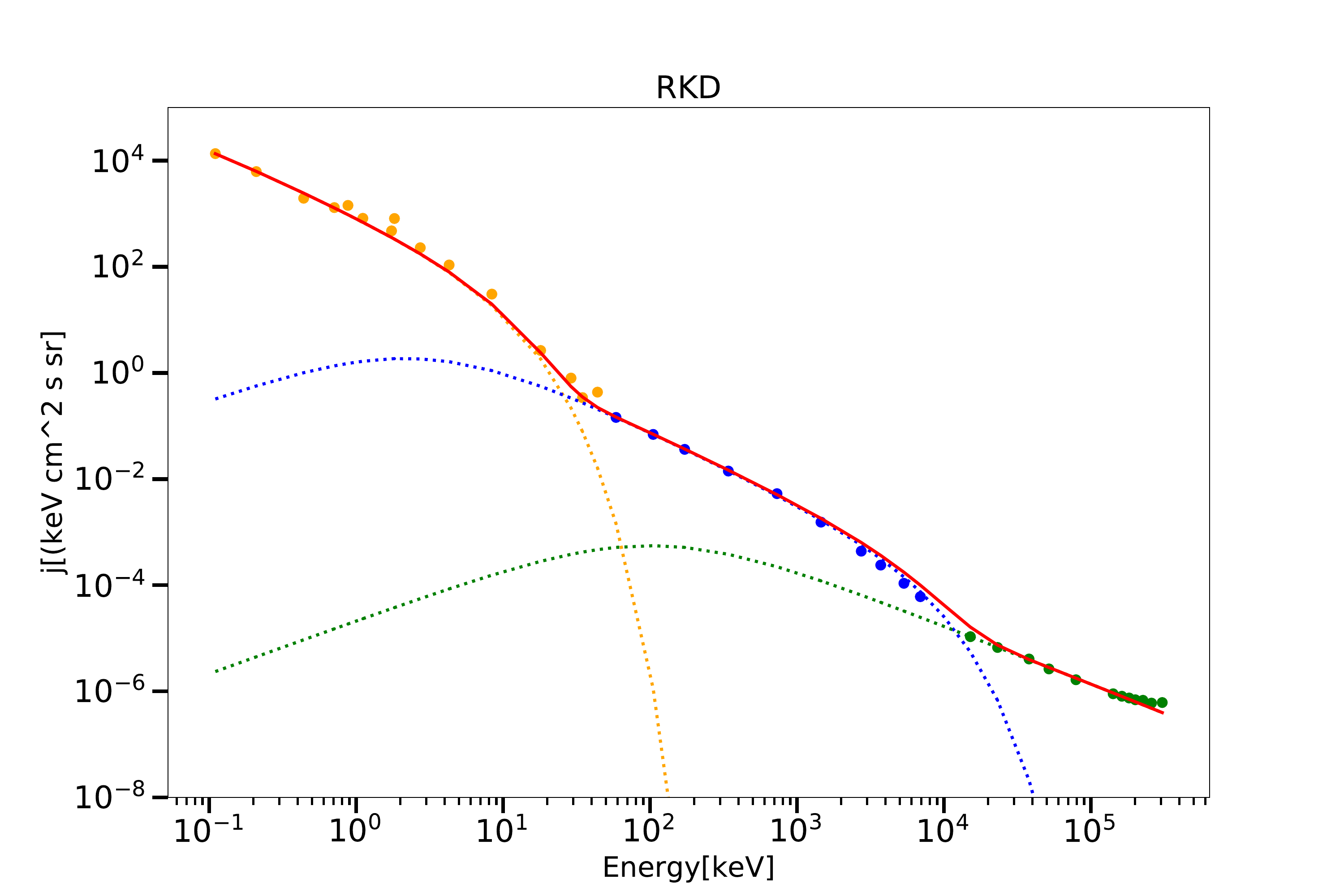}
  \includegraphics[width =0.95\columnwidth]{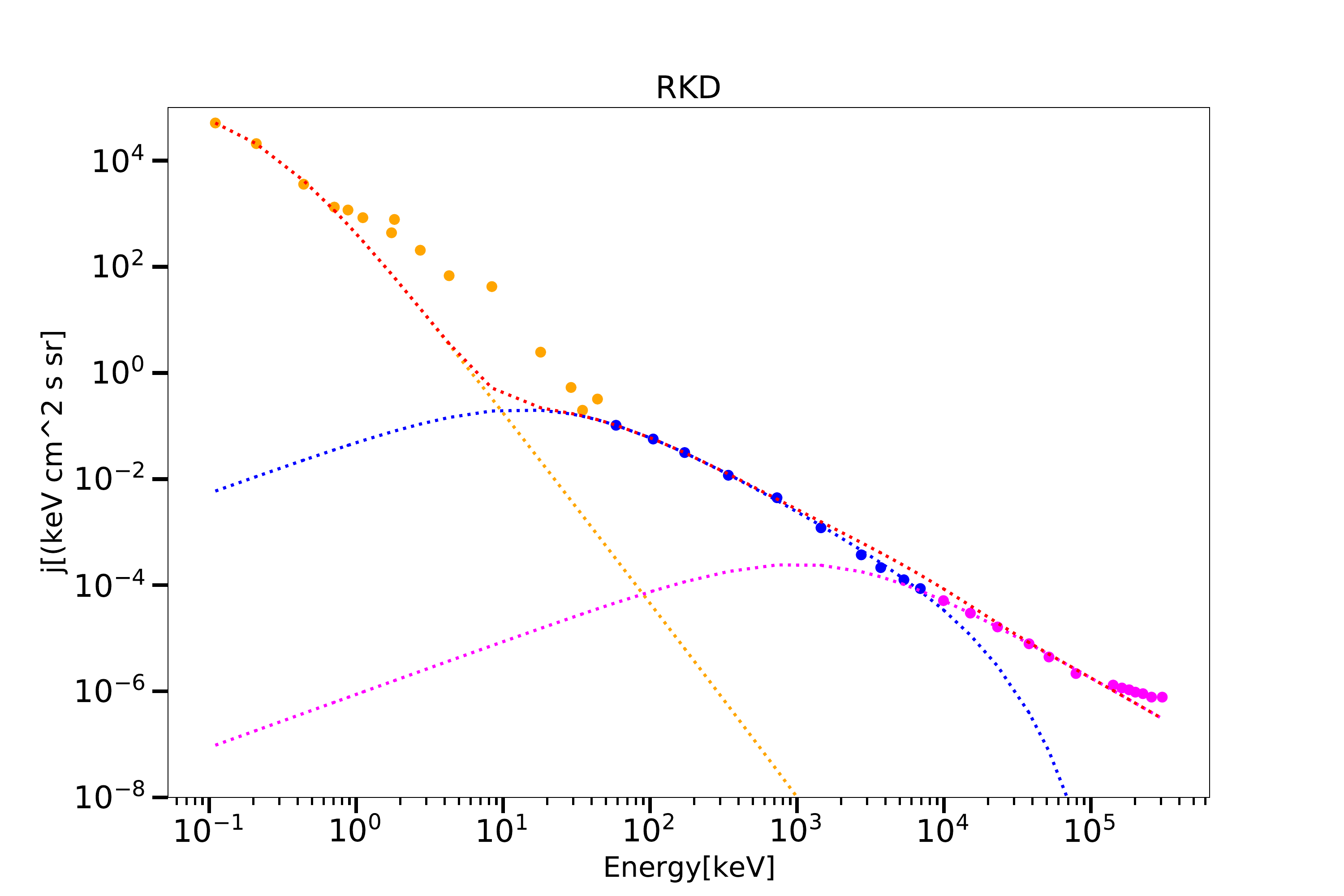}\\
  \includegraphics[width =0.95\columnwidth]{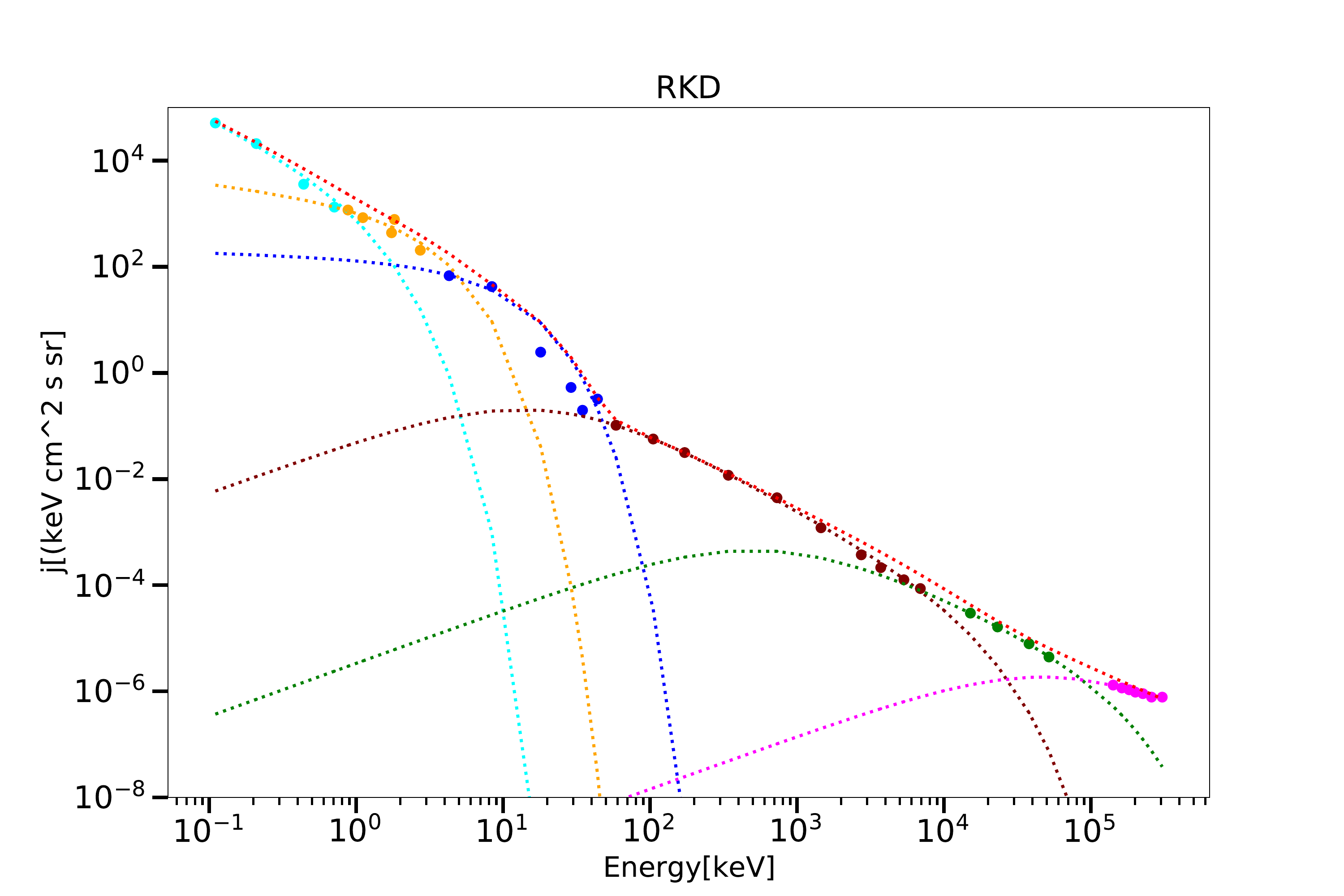}
  \includegraphics[width =0.95\columnwidth]{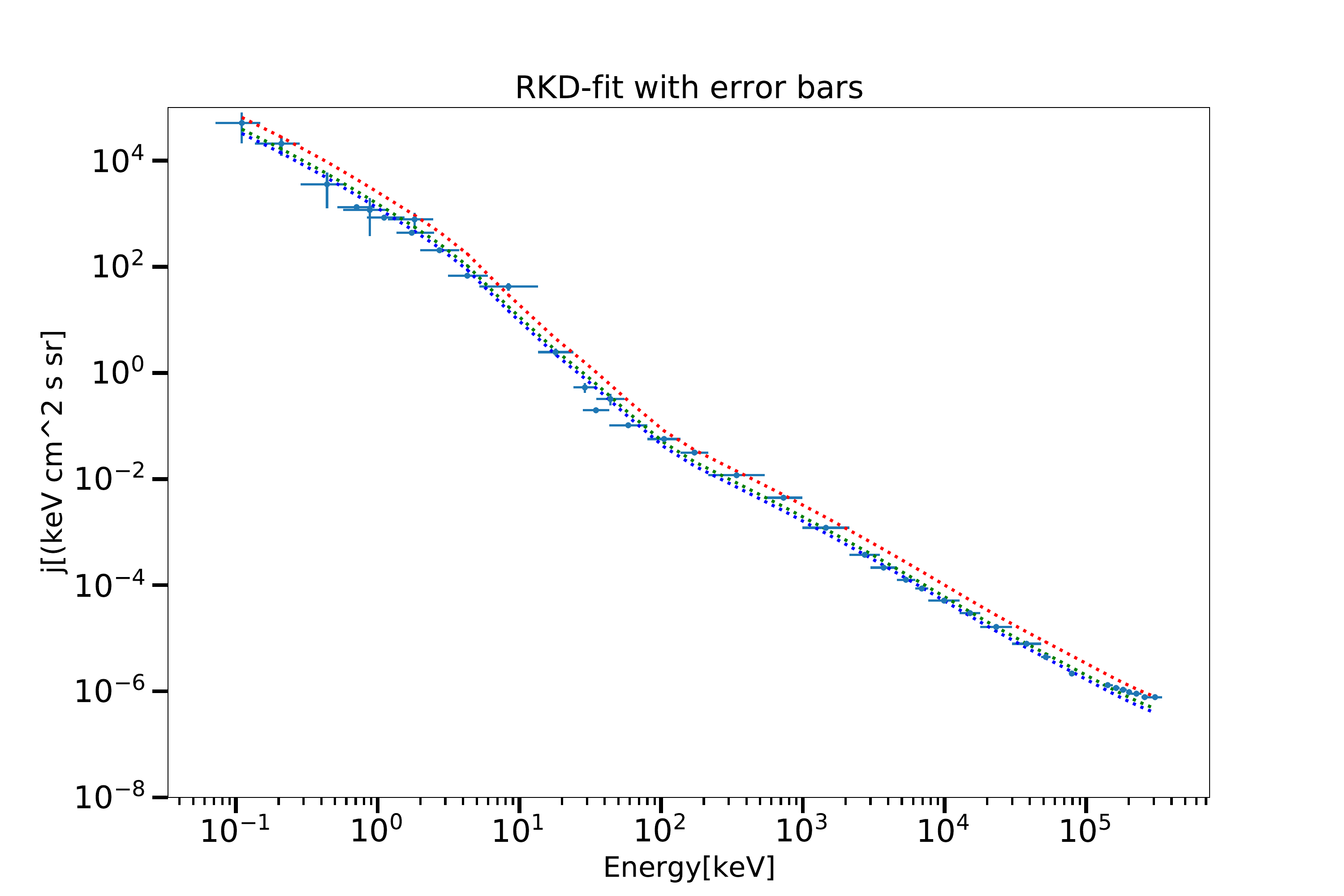}
  \caption{Upper left panel: Fit with three RKD distribution functions for the period 2009 to 2012. The yellow dotted line is the fit to the yellow data points, blue dotted line that to the blue marked data points and the green dotted line to green data points. The red line is the sum of the total distribution $f_t=\sum\limits_j f_f$. Upper right panel: Fit with three RKD distribution functions for the period 2013 to 2016. As one can see this fit is not good. Lower left panel: adaption to the data of period 2013 to 2016 with six RKDs. Lower right panel: the number densities as resulting from the 'fit' and after multiplication by a factor 0.6 (green dots) and a factor 0.5 (blue dots), which improves the total distribution function. The curves for the factor 0.5 and 0.6 are almost identical. Also the error bars for the data are shown. (See text for more discussion). \label{fig:rkd}}
\end{figure*}

We model the particle energy spectra shown in  Fig.~\ref{fig:rkd} with a combination of isotropic regularised $\kappa$-distributions (RKD). By eye-inspection we identify a couple of "cutoffs", that is energies where a power law is cutoff by the corresponding distribution function. Our analysis reveals the need for three RKD distributions to fit the period from 2009-2012 and six RKDs for the period  2013-2016. Before we go into the discussion of the "fitting" (see section~\ref{sec:fit} below) we present the RKD and the corresponding differential flux in the next section.

\subsection{The regularized distribution}\label{sec:distribution}
The isotropic regularized RKD was discussed in detail in \citet{Scherer-etal-2020}. For the convenience of the reader, we repeat it here. The RKD is given by :
\begin{align}\label{eq:rkd}
  f(v) = \frac{n_{0}}{\sqrt{(\pi^{3})} \kappa^{\thalf}\Theta^{3} U\left(\thalf,\thalf-\kappa,\xi^{2}\kappa\right)}
  \left(1+\frac{v^{2}}{\kappa\Theta^{2}}\right)^{-\kappa-1}
  e^{-\xi^{2}\frac{v^{2}}{\Theta^{2}}} 
\end{align}
where $v$ is the speed, $\Theta$ a normalisation speed, $n_{0}$ the
number density, $\kappa>0$ a
real number, and $\frac{\Theta}{c}<\xi<1$ a cutoff parameter, and $U(a,b,z)$ is the Kummer-U function \citep[e,g,\ ][]{Abramowitz-Stegun-1970}
The pressure is trace of the second order moment tensor, which for the isotrpic RKD yields: 
\begin{align}\label{eq:press}
    P &= P_{11}\\
    P_{RKD} & = \half m_p n_0 \Theta^2 
    \frac{U\left(\frac{5}{2},\frac{5}{2}-\kappa,\xi^2\kappa\right)}
    {U\left(\thalf,\thalf-\kappa,\xi^2\kappa\right)}\\
    P_{SKD} & = \half m_p n_0 \Theta^2 \frac{\kappa}{\kappa-\thalf}
\end{align}
We define the total RKD $f_t$ as:
\begin{align}
    f_t = \sum\limits_{i = 1}^n f_i
\end{align}
with, in our cases, $n=3$ or $n=6$. Thus, we have to put an index to $n_0, \Theta, \kappa,\xi$ in Eqs.~\ref{eq:rkd} and~\ref{eq:press}.The total pressure is the sum all partial pressures  $P_t= \sum\limits_i P_i$, as well as the total number density $n_{0,t} = \sum\limits_i n_{0,i}$.

The temperature for the partial pressures is given by the ideal gas law $T_i=P_i/n_{0,i}/k_{Bol}$, with the corresponding number density $n_{0,i}$ and the Boltzmann constant $k_{Bol}$. The total temperature is then:
\begin{align}\label{eq:totTemp}
    T = \frac{1}{n_{t}}\sum\limits_j n_j T_j
\end{align}

To get the differential flux, we replace the velocity by $v_i=\sqrt{2 E_i/m_i}$
where $E_i$ and $m_i$ are the energy and mass for the i'th RKD. To save writing we omit the indices and get
\begin{align}
  f(v(E))=&f(E) =\\
  &\frac{n_{0} m_{p}^{\thalf}}{(2\pi)^{\thalf} \kappa^{\thalf}E_{0}^{\thalf} U\left(\thalf,\thalf-\kappa,\xi^{2}\kappa\right)}
  \left(1+\frac{E}{E_{0}\kappa}\right)^{-\kappa-1}
  e^{-\xi^{2}\frac{E}{E_{0}}} 
\end{align}
With $E_{0}= m_{p}\Theta^{2} /2$ and $E=m_{p}v^{2}/2$ we
find for the differential flux $j= 2E/m_{p}^{2}f(v(E))$  in SI units (formula 6 from
\citep{Fahr-etal-2017} (replacing the electron mass by the proton
mass $m_e\to m_p\equiv m$:
\begin{align}
  j(E) &=
         \frac{n_{0}E}{\sqrt{2\pi^{3}m_{p}}E_{0}^{\thalf}\kappa^{\thalf}U\left(\thalf,\thalf-\kappa,\xi^{2}\kappa\right)}\left(1+\frac{E}{\kappa
         E_{0}}\right)^{-\kappa-1}
  e^{-\xi^{2}\frac{E}{E_{0}}} 
\end{align}
When we measure $E,E0$ in keV and $n$ in $cm^{-3}$ 
\begin{align}
  j(E) &= 3.93\cdot 10^6
         \frac{n_{0}E}{E_{0}^{\thalf}\kappa^{\thalf}U\left(\thalf,\thalf-\kappa,\xi^{2}\kappa\right)}\left(1+\frac{E}{E_{0}\kappa}\right)^{-\kappa-1}
  e^{-\xi^{2}\frac{E}{E_{0}}} 
\end{align}
where $j$ is now in [1/(\si{keV.s.cm^{2}.sr})]

As will be discussed in this study, while the SKD would not provide an adequate approximation, we introduce the differential flux of the SKD by \cite{Fahr-etal-2017} as:
\begin{eqnarray} 
j(E) = \frac{2 n_e E}{\pi^{3/2} m_e^2 U_\perp^3}
       \frac{\Gamma(\kappa+1)}{\kappa^{3/2}\Gamma(\kappa-1/2)} 
       \left[ 1+\frac{2 E}{\kappa E_0}\right]^{-\kappa-1}
\end{eqnarray} 
Here $U_\perp$ corresponds to the normalisations speed and $\Gamma(z)$ is the Gamma-function \citep[e.g.\ ][]{Abramowitz-Stegun-1970}.
In both cases (RKD and SKD), for large $E$ the distribution function follow
a power law form in energy of $\approx E^{-\kappa}$. For energies $E>E_0$ we can approximate the $\kappa$ part of the RKD 
\begin{align}
    j(E) & \propto E^{-\kappa}e^{-\xi^{2}\frac{E}{E_{0}}}    
\end{align}
(for the SKD without the exponential term). The exponential term approximates unity 
as long as the argument is (much) less than one ($\xi^2 E/E_0 <1$). 

For the total distribution function we thus get piecewise a power law for the distributions $f_i$. For the RKD we have a cutoff, which allow us to plot the differential flux for the total distribution, but not for the SKD, because there the power law of the first partial distribution dominates (see Fig.~\ref{fig:skd} and discussion below).
\begin{table*}[t!]
 \begin{tabular}
{l|@{}S[table-format = 1.2e1,table-column-width = 2cm]
@{}@{}S[table-format = 1.2,table-column-width = 1cm]
@{}@{}S[table-format = 1.2e1,table-column-width = 2cm]
@{}@{}S[table-format = 1.2e1,table-column-width = 2cm]
@{}@{}S[table-format = 1.2e1,table-column-width = 2cm]
@{}@{}S[table-format = 1.2e1,table-column-width = 2cm]
@{}@{}S[table-format = 1.2e1,table-column-width = 2cm]
@{}}
     & \multicolumn{1}{c}{$n_{0}$[cm$^{-3}$]} & 
     \multicolumn{1}{c}{$\kappa$} & 
     \multicolumn{1}{c}{E$_0$[keV]} & 
     \multicolumn{1}{c}{$\xi$} & 
     \multicolumn{1}{c}{$\Theta$[km/s]} & 
     \multicolumn{1}{c}{P[pPa]} &
     \multicolumn{1}{c}{T[K]}\\
$f_{1}$ &1.04e-2 & 1.29 & 6.51e-03 & 3.00e-02 & 3.53e+01 & 1.22e-01 & 8.53e+04 \\
$f_{2}$ &3.71e-6 & 1.34 & 2.15e+00 & 2.02e-02 & 6.42e+02 & 4.89e-05 & 9.56e+04  \\
$f_{3}$ &9.08e-9 & 1.09 & 1.01e+02 & 2.80e-03 & 4.39e+03 & 1.34e-06 & 1.07e+06 \\
total &1.04e-2 & {-}  & {-} & {-} & {-} & 1.22e-01 & 8.53e+04
\end{tabular}
  \caption{Parameter for the three fitted RKD- distributions and the total number density, pressure and temperature.The last line, indicated by total,  gives the sum of the  number density and pressures, as well as the temperature calculate by Eq.~\ref{eq:totTemp}.\label{tab:rkd}}
\end{table*}
 \begin{table*}[t!]
 \begin{tabular}
{l|
@{}S[table-format = 1.2e1,table-column-width = 2cm]
@{}S[table-format = 1.2e1,table-column-width = 2cm]
@{}@{}S[table-format = 1.2,table-column-width = 1cm]
@{}@{}S[table-format = 1.2e1,table-column-width = 2cm]
@{}@{}S[table-format = 1.2e1,table-column-width = 2cm]
@{}@{}S[table-format = 1.2e1,table-column-width = 2cm]
@{}@{}S[table-format = 1.2e1,table-column-width = 2cm]
@{}@{}S[table-format = 1.2e1,table-column-width = 2cm]
@{}@{}S[table-format = 1.2e1,table-column-width = 2cm]
@{}}
     & 
     \multicolumn{1}{c}{$n_{0}$[cm$^{-3}$]} & 
      \multicolumn{1}{c}{$0.6n_{0}$[cm$^{-3}$]} & 
     \multicolumn{1}{c}{$\kappa$} & 
     \multicolumn{1}{c}{E$_0$[keV]} & 
     \multicolumn{1}{c}{$\xi$} & 
     \multicolumn{1}{c}{$\Theta$[km/s]} & 
     \multicolumn{1}{c}{P[pPa]} &
     \multicolumn{1}{c}{0.6 P[pPa]} &
     \multicolumn{1}{c}{T[K]}\\  
$f_{1}$ &  2.00e-01 & 1.20e-01 & 1.35 & 1.00e-03 & 3.80e-02 & 1.38e+01 & 2.78e-01 & 1.67e-01 & 1.01e+04 \\
$f_{2}$ &  2.09e-03 & 1.25e-03 & 0.34 & 2.50e-03 & 3.67e-02 & 2.19e+01 & 1.02e-01 & 6.09e-02 & 3.52e+05 \\
$f_{3}$ &  3.15e-03 & 1.89e-03 & 1.73 & 4.69e-02 & 3.48e-02 & 9.48e+01 & 1.13e-01 & 6.77e-02 & 2.6e+05 \\
$f_{4}$ &  9.09e-07 & 5.46e-07 & 1.54 & 1.33e+01 & 3.37e-02 & 1.6e+03 & 1.24e-02 & 7.43e-03 & 9.86e+07 \\
$f_{5}$ &  1.35e-08 & 8.1e-09 & 1.29 & 5.02e+02 & 6.59e-02 & 9.81e+03 & 7.42e-03 & 4.45e-03 & 3.98e+09 \\
$f_{6}$ &  1.81e-09 & 1.09e-09 & 0.76 & 1.00e+03 & 1.00e-04 & 1.38e+04 & 5.48e+01 & 3.29e+01 & 2.19e+14 \\
total &  2.05e-01 & 1.23e-01 & {-} & {-} & {-} & {-} & 5.53e+01 & 3.32e+01 & 1.95e+06\\
total-$f_{6}$ &  2.05e-01 & 1.23e-01 & {-} & {-} & {-} & {-} & 5.13e-01 & 3.08e+01 & 1.95e+04
  \end{tabular}
  \caption{Similar to Table \ref{tab:rkd} For six distribution functions as shown in Fig.~ \ref{tab:rkd3}\label{tab:rkd3}}
  \end{table*}

\subsubsection{The "fit" procedure}\label{sec:fit}

The measurements shown in Fig.~\ref{fig:rkd} contain a total of 38 data points representing the differential flux as a function of energy, for both time periods considered.
We need to fit four parameters for each partial distribution function $f_i$. Thus, a fit with three distributions results in a total of 12 free parameters, whereas a six distribution fit results in 24 free parameter. Especially in the latter case, we have partial distributions, which can be "fitted" only for four points. Therefore, we refrain from naming it a "fit" to the data, but rather the optimum solution of the corresponding equation system. Thus the errors for the fit procedures are not well defined or may be large because of the small number of points. Nevertheless, we approximated the differential fluxes for both periods 

The period 2009-2012, shown in the upper left panel of Fig.~\ref{fig:rkd} and Tab.~\ref{tab:rkd}, requires the use of three distribution functions. The total fitted curve is given by the red line while the the partial fits are colored coded ($f_1$ in gold, $f_2$ in blue and $f_3$ in green). The differential flux for the total distribution fits the entire spectrum quite well. The numbers given in Tab.~\ref{tab:rkd} show also that the the $\kappa$-value is always below 1.5 which is the lower limit for the SKD. 

\begin{figure}[t!]
  \centering
  \includegraphics[width =\columnwidth]{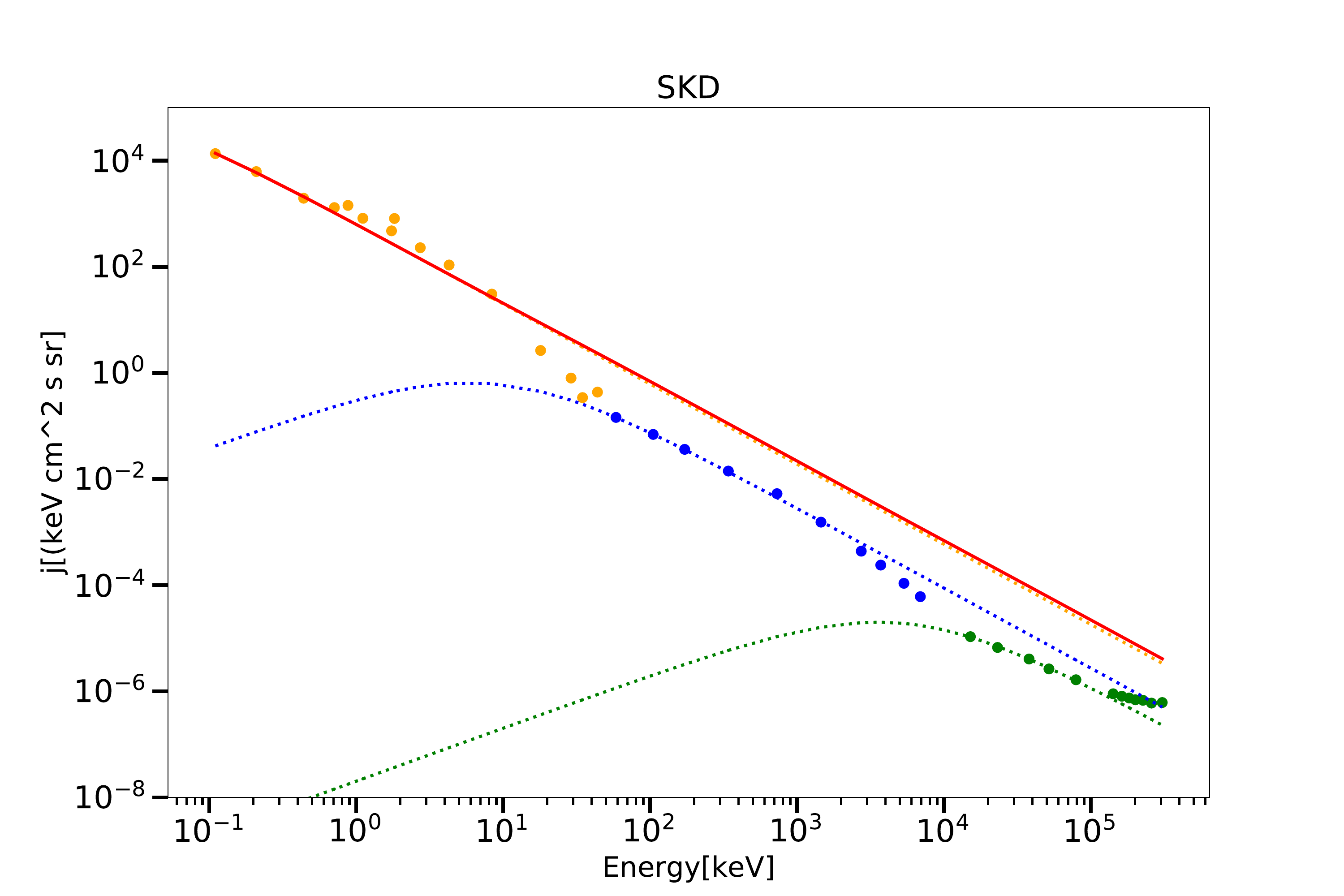}
  \caption{Fit with three SDK distribution functions. Same structure as in Fig.~\ref{fig:rkd}. The red line is the sum of all fitted distributions, which is nearly identically to the yellow line (covered by the red). See text for further discussion.\label{fig:skd}} 
\end{figure}

Nevertheless, we fitted three SKD's to the differential flux, as shown in Fig.~\ref{fig:skd} and Tab.~\ref{tab:skd}. Because there is no cutoff the spectrum is dominated by the the first SKD (gold color) moreover, it can be seen that the cutoffs are not well fitted, and that the $\kappa$-value is always 1.51 which is the lower boundary for the fitting procedure. Comparing the results between Tables~\ref{tab:rkd} and ~\ref{tab:skd} we notice that the total pressure and temperature for the SKD fit is larger by a factor $\sim$100. This is caused by the fact that the pressure close to the lower boundary for the SKD is not well defined (it increases quickly to infinity). 
The other reason for that behavior can be seen in Fig.~\ref{fig:skd} where the fit to the first distribution function $f_1$ covers the entire spectrum, and the area under this curve is much larger than that under the RKD $f_1$ curve for the same spectrum. Thus, one has to cut the distribution function artificially at the cutoff points in the spectra as discussed in \citet{Scherer-etal-2018,Scherer-etal-2019}. Then still the points around the cutoff remain which can not be fitted with a truncated SKD. Thus one needs to apply a RKD to get the observed cutoffs correctly, and to get more reliable pressures (temperatures).
Thus we do no longer discuss the SKD.

The energy spectra for the second period (2013-2016) could not be fitted with three $\kappa$-distributions, but required the use of six RKDs. The result is shown in the lower left panel of Fig.~\ref{fig:rkd} and in Tab.~\ref{tab:rkd3}. One can see from  Tab.~\ref{tab:rkd3} that the $\kappa$-values scatter between $\kappa=0.34$ to $\kappa=1.73$. Also the differential spectrum for the total fit is a somewhat higher than the data, which can be healed by multiplying the the sum of all RKDs by a factor 0.6. The result is shown in left lower panel of Fig.~\ref{fig:rkd} together with the error bars of the data points. Thus, we have to multiply all the number densities by a factor  0.6, as well as the pressure. The temperatures remain unchanged because $T\propto P/n$. Interestingly, the particle pressure that results from our analysis ($\sim$0.308 pPa) is only somewhat higher than the overall (isotropic) pressure in the IHS that was calculated directly from these measurements for the same time period \citep[][$\sim$0.251 pPa]{Dialynas-etal-2020}, a fact that further supports the relative success of our model and the need for the multiplying factor of 0.6. The overestimation of the particle pressure in our model by $\sim$20\% is a result of a slight overestimation of the particle fluxes in the $\sim$0.5-100 keV part of the energy spectra that carry a substantial amount of the total particle pressure inside the IHS. A multiplying factor of 0.5 provides a better representation of this part of the distribution (blue dotted line in the lower right panel of Figure~\ref{fig:rkd}) and corresponds to a total pressure of $\sim$0.256 pPa, that is in agreement with the \cite{Dialynas-etal-2020} calculated pressure. Furthermore, a total effective pressure of $\sim$0.267$\pm$0.55 pPa was calculated from \cite{rankin2019} using data-driven models and observations from IBEX.  

The choice of the the factor is quite arbitrarily and it is also possible, that for different parts of the adaption different factors can be used. Because these factors can not be accessed by the a fit, thus we have only given the number densities and pressures in Tab. 2 multiplied with the factor  0.6. One can easily change these values using the original number densities and pressures. The number densities are directly proportional to the distribution functions, and therefore also the pressure. The temperatures and the fitted energies $E_0$ (or corresponding normalization speeds $\Theta$), $\kappa$-values as well as the cutoff parameters are not affected by such a factor.

\section{Results}\label{sec:results}
The results for the RKD fit are shown in Fig~\ref{fig:rkd} and Tables~\ref{tab:rkd} and~\ref{tab:rkd3}. In the tables the fitted values for the number densities $n_{0,i}$ the characteristic energies $E_{0,i}$, the kappa value $\kappa_{i}$ and the cutoff parameter $\xi_{i}$ are is shown for the distributions in mind. That are for the period 2009-2012 (Tab.~\ref{eq:rkd}) three RKDs $f_{i}, i\in\{1,2,3\}$ and six RKDs $f_{i},i\in\{1,\ldots,6\}$  for the period 2013-2016. Also in the tables the derived parameter for the pressure, the temperature and characteristic speeds $\Theta_{0,i} = \sqrt{2E_{0,i}/m} $ are shown. In Table~\ref{tab:rkd3} additionally the values for the number densities and
pressures reduce by a factor  0.6 (0.5) are shown (see discussion above).

\subsection{The period from 2009 to 2012}
We concentrate the discussion first on the period 2009-2012 shown in Fig.~\ref{eq:rkd} and Table~\ref{tab:rkd}. From the latter table we see immediately, that the $\kappa$-value is always below 1.5 and thus can only be fitted by a RKD, and not and SKD. We also acknowledge that this time period corresponds roughly to the solar minimum of SC24, where the sunspot numbers on the sun and the solar wind pressure at $\sim$1 AU bottomed out at around 2010, presenting a maximum between 2014-2015 \citep[see also the discussion in][]{Dialynas-etal-2020}. We note that the slow solar wind needs about 1 year to reach the TS, while the fast solar wind takes only 0.5 years. Beyond the TS, the solar wind ions are transported through the IHS with radial velocities of $<$100 km/s that progressively decrease to zero, especially close to the HP \citep[e.g.][]{decker2012}. Note that the general morphology of the HP between the two voyager crossings was very similar, but there is a dramatic contrast between the plasma flow regime between the V1 encounter north of the heliographic equatorial plane \citep[stagnation region with tangential flow in -T direction]{krimigis2011} and V2 on the south \citep[transition region with the tangential flow in +T direction]{krimigis2019}. 

When we concentrate on the characteristic speeds (energies) we see that the three distinct RKDs have characteristic speeds which differ by orders of magnitude. That of the first RKD with the highest number density ($n_1\approx 0.01$\,\si{\# cm^{-3}}) has a characteristic speed of 35\,km/s, which can be attributed to lie within the IHS. The measurements from the plasma science experiment (PLS) on Voyager 2 beyond the TS \citep{richardson2014} over the 2009-2013 time period are consistent with convected Maxwellian distributions with typical thermal speeds of $\sim$30 km/s (with a spread between $\sim$20-50 km/s). In addition, the measured plasma temperature from V2/PLS over the 2009-2013 time period varies within $\sim40$-$90\, \si{kK}$ \citep{richardson2015}, which implies that the derived temperatures of $T_1 \approx 85\,\si{kK}$ fit in to that scenario. 

For the second RKD we have a characteristic speed of $\approx$650\,km/s and a temperature of $\approx$100\,\si{kK} which can be attributed to the termination shock particles and accelerated PUIs \citep{Chalov-Fahr-2000}. These authors  modeled the shock drift acceleration at the TS and found two populations of PUIs: A transmitted component, where the PUIs are transported through the TS and have a cuttoff around 10\,\si{keV} and a multiple reflected population with a cutoff around 500\,\si{keV} (they called it valley). These cutoffs coincide with the one found in the data and fitted by the RKD. The corresponding number density for the higher energetic particles (the transmitted PUIs) are quite low $n_1\approx 4\cdot 10^{-6}$\,\si{\# cm^{-3}}. 

Although highly sophisticated models that consider the heating and acceleration of PUIs at the TS \citep[e.g.][]{Giacalone-etal-2021} are constrained by the Voyager/LECP measurements, their inability to account for the measured fluxes of ENAs over the $\sim$0.5 to $\sim$55 keV energies \citep[e.g.][]{gkioulidou2022} implies that this population of particles is consistent with PUIs that are accelerated (or accelerated additionally) inside the IHS, i.e. downstream of the TS, and not necessarily at the TS. A recent analysis by \cite{Zirnstein-etal-2021} showed that the measured IBEX spectrum could not be obtained with protons being accelerated at the TS, unless using a ten times larger turbulence power ratio at the shock foot, than what was observed by Voyager \citep[e.g.][]{burlaga2008}. 

The third RKD has even lower number densities $n_3\approx 1\cdot 10^{-8}$\,\si{\# cm^{-3}} and characteristic speeds of 4440\,\si{km/s} and temperatures of $T_3\approx 1\,\si{MK}$. A close inspection of the  data points at the highest energies, show that they start to increase presenting a noticeable "hardening break" at $\sim$10 MeV (Fig. \ref{fig:rkd}), which is an indication of a contribution of galactic (GCR) and/or anomalous cosmic rays (ACR) \citep[see e.g.\ ][] {cummings2016,stone2019,Bischoff-etal-2019}. Earlier analyses attributed this feature in the spectra to a local disturbance at V2, that was most likely caused by a global merged interaction region (GMIR) passing through the IHS \citep{rankin2019}. However, as explained in \cite{Dialynas-etal-2020}, this is a result of the unfolding of the ACR spectrum inside the IHS (up to at least the year 2011), which was not the expected power law at the TS during the V2 crossing.

The sources of ACRs, generated from solar wind PUIs (that are the dominant energy source in the IHS) are different from those of GCRs. Previous analyses comparing the ACR energy spectra with the estimated flux of PUIs at the termination shock revealed a mass-dependent acceleration that favors heavier ions \citep[][ and references therein]{cummings2007} and the analysis of \cite{cummings2019} supported the idea of ACRs being accelerated back towards the flank or tail of the TS. However, other analyses have shown that different processes, such as magnetic reconnection \citep{drake2010, zank2015, Zhao2019} or the so-called pump mechanism \citep{fisk2012} can also potentially accelerate ACRs inside the IHS, whereas these particles are influenced by interstellar PUIs \citep{hill2020}. 

At higher energies, the spectra beyond about 100 MeV have most likely contributions from GCR \citep{Dialynas-etal-2020}, and as shown in our analysis they cannot be fitted with a power law or a $\kappa$-distribution. This is indicated by the very low $\kappa=0.76$ value, which is consistent with the fact that the interstellar GCR spectrum at low energies is almost flat \citep[e.g.][]{stone2019,Bischoff-etal-2019}. While the RKD does not capture the high energy tail that is modulated by GCRs, we note that a sufficient approximation of the $>$300 MeV GCR spectra in interplanetary space was based on a force-field approximation \citep[e.g.][and references therein]{roussos2020}, whereas the measured GCR spectra upstream at the HP together with estimated spectra in the VLISM using a variety of different models (e.g. \cite{engelmann1990}; leaky-box model and \cite{vladimirov2011}; GALPROP model \citep{Bischoff-etal-2019}) are shown in \cite{cummings2016}.

\subsection{The period from 2013 to 2016}
The upper right panel of Fig.~\ref{fig:rkd} provides a fit with three distribution functions, which is clearly inadequate to describe the measured spectra, and undershoots substantially the proton fluxes, especially within the energy range of $\sim$1 to 55 keV. We, thus, consider an adaption with six RKDs, shown in the lower left panel of Fig.~\ref{fig:rkd}. Because the sum of the adaptions (red dotted line shown in the lower right panel of Fig.~\ref{fig:rkd}) is to large by a factor of $\sim$0.6 (or $\sim0.5$; see discussion in Section~\ref{sec:models}), we multiply the number densities with this number, toward producing a better approximation of the measurements (green dotted line for the factor 0.6 and blue dotted line for the the factor 0.5). As explained earlier, the approach of using a set of six distribution functions cannot be called a fit to the data, but rather a solution of the corresponding equation system. Thus, this is rather an adaption to the data with six distribution function than a fit.

The reason for this behavior is not surprising, but is due to the fact that the ascending phase of solar cycle 24 (2013-2016) has not yet filled the entire IHS, and thus we have a combination of the proton fluxes from the previous cycle 23 and the ascending cycle 24. This can be seen in the first three distribution functions which have slightly different characteristic speeds (($\Theta_{1,2,3} \approx(14,22, 95]$\,\si{km/s}). The same holds true for the fourth and fifth distribution ($\Theta_{4,5} \approx(1.6,9.8)$\,\si{Mm/s}, while the sixth distribution has a characteristic speed of $\Theta\approx 14$\,\si{Mm/s} and resembles most probably a mixture of ACRs and GCRs. 

Notably, the $<$28 keV measurements in the spectra shown in Fig.~\ref{fig:rkd} present the largest discrepancy with the three $\kappa$-distribution approach and correspond to measured ENAs (IBEX-Hi and INCA) that have been converted to protons. The ENA fluxes at the IBEX energies generally respond to the SW pressure changes over the solar cycle somewhat slower than the ENAs corresponding to the energies covered by INCA. For example, unlike the $>$5.2 keV ENA measurements from INCA and in-situ ions from LECP \citep{dialynas2017}, the $<$6 keV ENAs do not present a local minimum within the year 2013, and require more time to respond to the min-to-max pressure changes over the SC than higher-energy ENAs and in-situ ions \citep[e.g.][]{mccomas2020}. However, it should be noted that the response of the $\sim$4.29 keV ENA IBEX channel exhibits the shortest time delay and largest flux variation \citep{reisenfeld2016, Zirnstein2018, mccomas2020} in concert with the $>$5.2 keV ENAs from INCA \citep{dialynas2017, dialynas2019} that respond within $\sim$2-3 years, on average. A response to a large solar wind intensification in the upwind hemisphere (but not downwind), with a time delay of
$\sim$2-3 yr, especially of the higher energy ($\sim$4.29 keV) ENAs, has
been previously reported in \cite{mccomas2018}. 

The differences between the two periods under discussion can be identified via the recorded ICME/CME/SSS events at about \si{1 au}, considering the lists compiled by the George Mason University, Space Weather Lab (\url{http://solar.gmu.edu/heliophysics/index.php/The_ISEST_ICME\%5CCME_Lists}). Because these disturbances usually need 1-3 years to reach the IHS we can count the number of disturbances in the time period from 2007 to 2010 (or going only 1 year back between 2009 to 2011) to be about 31 (61) of such events. In the later period, from 2011 to 2014 (2012 to 2015), there are  more than 100 (100) events. From those numbers we assume that the state of the IHS between 2009 and 2011 is quite calm as the solar wind was quite calm at \si{1 au}, while it is highly dynamic in the latter period. Thus, although it is possible to fit the state of the IHS in the first period (2009 --2012), the latter time period (2013 -- 2016), where the maximum conditions of the solar cycle 24 reach the IHS, presents further complications that render such an attempt rather challenging, as explained in this section (see also Section \ref{sec:discussion}).

\section{Discussion}\label{sec:discussion}

\cite{Chalov-Fahr-2000} studied the particle energy spectra downstream of the TS focusing at an energy range of $10^{-1}$ to $10^4$ \si{keV}, while recent analyses simulated only parts of the ion spectrum. For example, \cite{Baliukin-etal-2022} provided a good approximation to the corresponding part of the spectrum (IBEX-Hi), whereas other authors compared the different types of shock acceleration (diffusive shock versus shock drift acceleration), see \citet{Zirnstein-etal-2021}, or simulate it locally \citep{Lembege-etal-2018,Giacalone-etal-2021,Wu-etal-2016}. A recent simulation of most parts of the ENA and PUI-spectrum can be found in \citet{Czechowski-etal-2020} which is based on the model of PUI acceleration at the TS proposed by \citet{Zank-etal-2010}.

Other authors concentrate more on simulating ENA spectra \citep{Galli-etal-2019,Zirnstein-etal-2021}, which show similar behavior as the fitted spectra, discussed above.

A common problem among most of these sophisticated modeling efforts that simulate the proton energy spectra, especially within the critical part of the distribution that corresponds to the $>$0.5 to 55 keV energies, which provides a significant amount of pressure inside the IHS, is that they substantially underestimate the measured fluxes. \cite{zirnstein2017} discuss the persistent discrepancy between their model and the 0.71-4.29 keV IBEX measurements of a factor of $\sim$2-3, as a possible result of a miss-estimation for the IHS thickness (a common issue among most of the advanced heliosphere models, \citep[e.g.][]{kleimann2022}), and/or a less constrained velocity diffusion of ions inside the IHS, and/or because of adopting less precise ion distribution function downstream of the TS. The model of \cite{Baliukin-etal-2022} employs a kinetic treatment of PUIs and produce simulated spectra that differ by a factor of only $\sim$ 1.5 from the measured data.

At higher energies, the ENA simulations from \cite{Czechowski-etal-2020} are lower by a factor of $>$4 from the $>$5.2 keV measured ENAs \citep{dialynas2017b} for both solar minimum and maximum conditions. Moreover, the simulated energy spectra in \cite{Czechowski-etal-2020} are substantially harder than the measured spectra. As explained in \cite{dialynas2019}, underestimating the flux and spectral slope of this part of the proton distribution leads to underestimating considerably the partial particle pressure inside the IHS, thus differing from obtaining realistic numbers for the pressure balance that forms our solar bubble. Recently, \cite{gkioulidou2022} provided hybrid simulations of ions downstream of the TS using the sophisticated \cite{Giacalone-etal-2021} model and showed an energy depended discrepancy between the 0.52-55 keV measured (IBEX-Hi and INCA) and simulated ENA fluxes, with the observations being persistently higher than the model, suggesting that further acceleration of PUIs occurs within the IHS. 

There is also an attempt to model the the PUI flux in the IHS by \citet{Fahr-etal-2016} using  a SKD. Some of these models have the problem that using a SKD limit the power index $\kappa$ to values above 1.5. While our fit show that the best values of $\kappa$ are below that value for all three fitted distributions: $\kappa_{1,2,3}\approx{1.3,1.3,1.1}$ see Tab.~\ref{tab:rkd}. 

The advantage of the RKD fit to the spectra is that we can estimate the macroscopic velocity moments (density and pressure) and assuming that the ideal gas law holds, we can estimate the temperatures. We point out that the higher order velocity moments $l$ of the RKD increase with $\kappa\to 0$ as $1/([l+1]\xi^l)$ \citep{Husidic-etal-2022}, and thus even the small $\kappa$ values, like $\kappa=0.34$ for the second adapted distribution in the period 2013 to 2016 has a finite pressure (temperature).

When we look at the total number densities we recognize, that they agree quite well with the standard assumption of the interstellar neutral gas densities \citep{dialynas2019, Dialynas-etal-2020, swaczyna2020} in 2009-2012, while those in 2013-2016 are slightly larger when we use the factor 0.6 for the adaption of the distribution functions to the spectra. Nevertheless, the latter period should be handled with care because we do not have a quasi-stationary state, but the solar wind speed is strongly varying between  high and low speed streams \citep{McComas-etal-2008}, while during the first period (approximately solar minimum) we have a quasi-stationary state of the heliosphere, because the high speed wind are now locate above \ang{35} while the low speed wind is locate below \ang{35} and the entire solar wind is quite smooth. During the latter period we have then also a quasi-stationary TS for which the shock drift acceleration \citep[e.g.\ ][]{Chalov-Fahr-2000} applies, while during the period 2013-2016 the TS moves in and out \citep[][and refernces therein]{Strumik-Ratkiewicz-2022}. 
Thus the TS moves over the declining phase of each solar cycle with some time delay because of the finite travel time of the solar wind. That is the difference between the two periods, in fact  SC24 has the weakest conditions from all known solar cycles. The minimum of SC23 was deep and very prolonged, which can account for a "quasi-stationary" condition inside the IHS \citep{McComas-etal-2013}.  Thus the dynamics of the change between SC23 and SC24 explain the differences between the 2009-2013 and 2013-2016 time period.

For the adaption of the spectrum in 2013-2016 the sum of the distributions functions is a little bit higher than that of adapted single distributions. The reason is that adaption fits the required points quite well, but the sum is slightly too large. The multiplication with all number densities by a factor 0.6 (0.5) gives better results. These values (number densities and pressures)  are indicated in Tab.~\ref{tab:rkd3} by the factor 0.6. Further we have calculated the total number densities, pressures and temperatures without the contribution of $f_6$, because that will give to high values. The reason is that it continues to high energies, where one has to take into account the modulation of ACRs and GCRs, which will increase towards 1\,\si{MeV} or 1\,\si{GeV} and then fall off exponentially as $\sim E^{-2.3}$ \citep{Bischoff-etal-2019} to $\sim E^{-2.7}$ \citep{fisk2012} toward the knee of the GCRs that occurs at $\sim$8$\times 10^{6}$ GeV.     

%

\begin{table*}[t!]
\begin{tabular}
{l|
@{}S[table-format = 1.2e1,table-column-width = 2cm]
@{}@{}S[table-format = 1.2,table-column-width = 1cm]
@{}@{}S[table-format = 1.2e1,table-column-width = 2cm]
@{}@{}S[table-format = 1.2e1,table-column-width = 2cm]
@{}@{}S[table-format = 1.2e1,table-column-width = 2cm]
@{}@{}S[table-format = 1.2e1,table-column-width = 2cm]
@{}@{}S[table-format = 1.2e1,table-column-width = 2cm]
@{}}
     & 
     \multicolumn{1}{c}{$n_{0}$[cm$^{-3}$]} & 
      \multicolumn{1}{c}{$\kappa$} & 
     \multicolumn{1}{c}{E$_0$[keV]} & 
    \multicolumn{1}{c}{$\Theta$[km/s]} & 
     \multicolumn{1}{c}{P[pPa]} &
     \multicolumn{1}{c}{T[K]}\\      
$f_{1}$ &  1.00e-02 & 1.51 & 8.45e-03 & 4.02e+01 & 1.02e+01 & 7.40e+06 \\
$f_{2}$ &  2.00e-06 & 1.51 & 5.93e+00 & 1.07e+03 & 1.43e+00 & 5.2e+09 \\
$f_{3}$ &  1.50e-09 & 1.51 & 3.55e+03 & 2.61e+04 & 1.53e-06 & 7.40e+06 \\
total &  1.00e-02 & {-} & {-} & {-} & 1.17e+01 & 8.44e+06
  \end{tabular}
  \caption{Parameter for the three fitted SKD- distributions, and the total number density, pressure and temperature.\label{tab:skd}}
\end{table*}

\section{Summary and Conclusion}\label{sec:summary}

We demonstrated that \st{one can fit} the 0.11~keV to 344~MeV composite proton spectra during the solar minimum period from 2009 to the end of 2012 can be fitted with three RKDs, namely
for the lowest energy range (RKD $f_1$) the transmitted PUIs and for the middle energy range (RKD $f_2)$ the multiply reflected PUIs, while the highest energies (RKD $f_3$) are most probably a mixture of the highest transmitted PUIs and modulated ACRs/GRCs. 

For the subsequent period 2013 to 2016, i.e.\ a period of solar maximum, a fit needs six rather than three distributions. The reason may be the highly dynamic IHS during that period. Our approach indicates that  there seems to be a mixture of transmitted particles from the minimum of cycle 23 and from the maximum of solar cycle 24 in the lowest and middle energy range. The highest energy range splits into high-energetic transmitted particles or pre-accelerated ACRs and a 
high-energy component of GCRs which should not be represented by an RKD. Overall, as discussed in Section~\ref{sec:discussion} our results are consistent with PUI and ACR shock accelerated particles that undergo additional acceleration inside the IHS.

Albeit the description of the dynamic state of the IHS with ion spectra that are averaged over 3 to 4 years can be potentially problematic, we stress that during the first period (2009 to 2012) the state of the solar wind was quite calm when it reached the TS and thus we could fit the average ion spectrum quite adequately, attributing it to the transmitted and reflected PUI population. In contrast to that, the highly dynamic state of the IHS during the ascending phase of SC24 toward solar maximum made the task of fitting the ion measurements during the second time period (2013 --2016) challenging. During that time, the state in the IHS seems to be quite chaotic. Nevertheless, our analysis for the subject time period resulted in a total pressure for the IHS that is comparable to the pressures provided by \cite{Dialynas-etal-2020}, who accurately obtained the magnitude of the magnetic field upstream at the HP by performing a pressure balance at the boundary. This indicates that our approach does in fact capture part of the dynamics of the IHS, as discussed in Section \ref{sec:results}.

Besides the fact that these data have been fitted with physics-based
distribution functions for the first time, the results represent another 
demonstration of the usefulness of RKDs, because equivalent fits with 
standard kappa distributions would not be possible.

\begin{acknowledgements}
      KD acknowledges the support from the NASA contracts NAS597271, NNX07AJ69G, and NNN06AA01C and subcontract at the Office for Space Research and Technology (Academy of Athens), together with useful discussions and collaboration with all SHIELD team members (NASA grant 18-DRIVE18\_2-0029, Our Heliospheric Shield, 80NSSC20K0603: \href{http://sites.bu.edu/shield-drive/}{http://sites.bu.edu/shield-drive/}) that made this work possible. KS, KD, HF and AG are grateful to join the ISSI workshop "The Heliosphere in the Local Interstellar Medium" and the useful discussions therein.
\end{acknowledgements}

\bibliographystyle{aa}
\bibliography{ena}

\end{document}